\documentclass[twocolumn,showpacs,preprintnumbers,tightenlines,amsmath,amssymb]{revtex4-1}

\usepackage[colorlinks,pdfstartview=FitH]{hyperref}
\hypersetup{linkcolor=blue,citecolor=blue,filecolor=black,urlcolor=blue}

\usepackage{amsmath,amssymb,bm}
\usepackage{graphicx}
\usepackage{amsfonts}
\usepackage{color}

\usepackage{amsfonts}
\usepackage{bbm}
\usepackage{latexsym}
\usepackage{slashed}

\begin{document}


\baselineskip=15pt \parskip=5pt

\vspace*{3em}

\title{Could the 21-cm absorption be explained by the dark matter suggested by $^8$Be transitions?}

\author{Lian-Bao Jia }
\email{jialb@mail.nankai.edu.cn}
\affiliation{School of Science, Southwest University of Science and Technology, Mianyang
621010, China}

\author{Xian-Jin Deng}
\affiliation{School of Science, Southwest University of Science and Technology, Mianyang
621010, China}

\author{Chang-Fu Liu}
\affiliation{School of Science, Southwest University of Science and Technology, Mianyang
621010, China}

\begin{abstract}

The stronger than expected 21-cm absorption was observed by EDGES recently, and another anomaly of $^8$Be transitions would be signatures of new interactions. These two issues may be related to each other, e.g., pseudoscalar $A$ mediated fermionic millicharged dark matter (DM), and the 21-cm absorption could be induced by photon mediated scattering between MeV millicharged DM and hydrogen. This will be explored in this paper. For fermionic millicharged DM $\bar{\chi} \chi$ with masses in a range of $2 m_A < 2 m_{\chi} < 3 m_A$, the p-wave annihilation $\bar{\chi} \chi \to A A$ would be dominant during DM freeze-out. The s-wave annihilation $\bar{\chi} \chi$ $\to A, \gamma $ $\to e^+ e^-$ is tolerant by constraints from CMB and the 21-cm absorption. The millicharged DM can evade constraints from direct detection experiments. The process of $K^+ \to \pi^+ \pi^0$ with the invisible decay $\pi^0 \to \bar{\chi} \chi$ could be employed to search for the millicharged DM, and future high intensity $K^+$ sources, such as NA62, will do the job.

\end{abstract}

\maketitle


\section{Introduction}

Recently, a stronger than expected absorption of the global 21-cm spectrum at a redshift of $z\sim$17 was reported by the EDGES Collaboration \cite{Bowman:2018yin}, with a significance of 3.8 $\sigma$. This anomaly
may be due to the hydrogen gas cooled by the photon mediated scattering with dark matter (DM) at the cosmic dawn, i.e., a small fraction about [DM mass (MeV)/10]$\times$0.115\%$-$0.4\% of DM carrying a millicharge $\eta e$ (with $\eta$ $\sim 10^{-4} - 10^{-6}$ and DM mass in a range of 10$-$35 MeV) \cite{Barkana:2018lgd,Munoz:2018pzp,Fialkov:2018xre,Xu:2018efh,Berlin:2018sjs,Barkana:2018qrx,Slatyer:2018aqg,Munoz:2018jwq,Mahdawi:2018euy,Boddy:2018wzy,Kovetz:2018zan}.
Moreover, other possible explanations about the anomaly, such as additional radiation background at a relevant low frequency, are considered in Refs. \cite{Feng:2018rje,Fraser:2018acy,Mirocha:2018cih,Pospelov:2018kdh,Moroi:2018vci}. Further exploration about the 21-cm spectrum during the dark ages \cite{Barkana:2016nyr,Hektor:2018qqw,Li:2018kzs,Hektor:2018lec} may probe more properties of DM.

Here the millicharged DM explanation is of our concern. To obtain the small fraction of millicharged DM, large DM annihilation cross sections caused by new interactions are required during DM freeze-out. In addition, the observations of the cosmic microwave background (CMB) at the recombination \cite{Ade:2015xua,Slatyer:2015jla} and the 21-cm absorption at the cosmic dawn \cite{Cheung:2018vww,Liu:2018uzy,DAmico:2018sxd} set constraints on DM annihilations with masses of tens of MeV. To evade these constraints, scenarios of DM annihilating into neutrinos \cite{Berlin:2018sjs}, or DM annihilations in p-wave \cite{Berlin:2018sjs,Jia:2018csj} during DM freeze-out are available.

Possible types of new interactions between millicharged DM and standard model (SM) particles are unclear. Recently, an indication of new interactions was observed in the invariant mass distributions of $e^+ e^-$ pairs produced in $^8$Be transitions \cite{Krasznahorkay:2015iga}, which cannot be explained within nuclear physics \cite{Krasznahorkay:2015iga,Zhang:2017zap}. A new vector boson $X$ being produced and quickly decaying via $X \to e^+ e^-$ was suggested to explain the anomaly, with the mass $m_X^{} \simeq$ 17 MeV. Possible vector/axial vector couplings of $X$ with SM fermions were analyzed in Refs. \cite{Feng:2016jff,Feng:2016ysn,Kozaczuk:2016nma,DelleRose:2017xil} (for more discussions, see e.g., Refs. \cite{Gu:2016ege,Liang:2016ffe,Fornal:2017msy,Banerjee:2018vgk,Jiang:2018uhs}), and the vector/axial vector $X$ portal DM particles were studied in Refs. \cite{Jia:2016uxs,Kitahara:2016zyb,Chen:2016tdz,Jia:2017iyc,Seto:2016pks}. In addition, a pseudoscalar $A$ with the mass about 17 MeV may also produce $^8$Be anomalous transitions \cite{Ellwanger:2016wfe}.

In the case that the MeV DM suggested by the $^8$Be transitions is millicharged, the new interaction portal DM may give an explanation on the EDGES observation. This is of our concern in this paper. For the vector $X$ portal millicharged DM \cite{Jia:2016uxs}, a large $X$-DM coupling is needed to obtain the small fraction of millicharged DM.

In this paper, we focus on the pseudoscalar $A$ mediated fermionic DM, which is millicharged. When DM is heavier than the pseudoscalar mediator, DM can annihilate both in s-wave and p-wave. It may be allowed by the constraints from the CMB and 21-cm absorption observations, and gives an alternative explanation about the 21-cm anomaly. These will be investigated in the following.

\section{Interactions and transitions}

The effective couplings of the pseudoscalar $A$ to SM quarks are taken in the form
\begin{eqnarray}
\mathcal{L}_{A q}= \xi_q \frac{m_q}{v} A \bar{q} i \gamma_5 q,
\end{eqnarray}
where the vacuum expectation value $v$ is $\sim$ 246 GeV. With the assumption of $m_d \sim 2 m_u \sim 2 \times 2.5$ MeV \cite{Cheng:2012qr} and $\xi_u = \xi_c = \xi_t$, $\xi_d = \xi_s = \xi_b$, to explain the $^8$Be anomaly, the values of $\xi_u + \xi_d \approx$ 0.6 and $\xi_e \gtrsim 4$ can be adopted \cite{Ellwanger:2016wfe}. In addition, the coupling parameter between a new pseudoscalar particle (with a mass $\sim$ 17 MeV) and electron is $\xi_e \gtrsim 115$ in Ref. \cite{Berlin:2018bsc} (referencing the E141 result \cite{Riordan:1987aw}). Furthermore, if $A$ couples to muon, it will be constrained by the muon $g-2$. The one-loop result of the pseudoscalar $A$ is \cite{Hektor:2015zba}
\begin{eqnarray}
a_{\mu}^A=  \frac{m_\mu^2 \xi_\mu^2}{8 \pi^2 v^2}  \kappa \int_0^1 \mathrm{d} x \frac{ - x^3}{1-x  +x^2 \kappa} ,
\end{eqnarray}
where $\kappa = m_\mu^2 / m_A^2 $. The recent result for the discrepancy between experiment and theory is about \cite{Patrignani:2016xqp,Keshavarzi:2018mgv,Borsanyi:2017zdw,Blum:2018mom}
\begin{eqnarray}
\Delta a_{\mu}=  a_{\mu}^{\mathrm{exp}} - a_{\mu}^{\mathrm{SM}} \simeq ( 2.7 \pm 0.7 ) \times 10^{-9} .
\end{eqnarray}
Suppose $A$'s contribution to the muon $g-2$ difference is $\lesssim 1 \times 10^{-9}$. For the case of Higgs-like couplings of $A$ to leptons, i.e., $\xi_{\mu} = \xi_e$, this will significantly enlarge the discrepancy. For the case of universal couplings of $A$ to electron and muon, i.e., $m_{\mu} \xi_{\mu} \sim m_e \xi_e$, we have $\xi_e \lesssim 196$.

The effective coupling of $A$ to the fermionic millicharged DM $\chi$ is taken as
\begin{eqnarray}
\mathcal{L}_A^{\mathrm{DM}}= \lambda A \bar{\chi} i \gamma_5 \chi .
\end{eqnarray}
For DM being heavier than $A$, DM can annihilate both in s-wave and p-wave. In addition, to avoid the s-wave annihilation $\bar{\chi} \chi \to A A A$ after DM freeze-out (see Appendix \ref{appendix:three-A} for more), a mass range of DM $2 m_A < 2 m_{\chi} < 3 m_A$ is considered.

Now we formulate the annihilations of millicharged DM $\bar{\chi} \chi$. The annihilation cross section of the p-wave process $\bar{\chi} \chi$ $\to A A$ is
\begin{eqnarray}
\sigma_0 v_r \simeq \frac{1}{2} \frac{\lambda^4  m_{\chi} ( s/4 - m_A^2 )^\frac{5}{2} }{12 \pi (s - 2 m_\chi^2)} \frac{(s - 4 m_\chi^2 )}{(m_A^2 - 2 m_{\chi}^2)^4 } ,
\end{eqnarray}
where $v_r$ is the relative velocity of the annihilating DM pair. The factor $\frac{1}{2}$ is for the required $\bar{\chi} \chi$ pair in DM annihilations. In the nonrelativistic limit, the total invariant mass squared $s$ is $s = $ $4 m_\chi^2 + m_\chi^2 v_r^2 + \mathcal{O} (v_r^4)$.

The s-wave processes of DM annihilations are mainly mediated by $A$ and $\gamma$. For the process $\bar{\chi} \chi$ $\to A$ $\to e^+ e^-$, the annihilation cross section is
\begin{eqnarray}
\sigma_1 v_r \simeq \frac{1}{2} \frac{\lambda^2  \xi_e^2 m_e^2/v^2 }{16 \pi (s - 2 m_\chi^2)} \frac{s^2}{(s -  m_{A}^2)^2}  .
\end{eqnarray}
For the DM mass of concern, the electron's mass is negligible compared with DM mass. For photon mediated transitions, the annihilation mode $\bar{\chi} \chi$ $\to \gamma $ $\to e^+ e^-$ is an s-wave process, which is suppressed by $\eta^2$. The corresponding annihilation cross section is
\begin{eqnarray}
\sigma_2 v_r \simeq \frac{1}{2} \frac{2 \pi \alpha^2 \eta^2 }{(s - 2 m_\chi^2)} .
\end{eqnarray}
Another s-wave process $\bar{\chi} \chi$ $\to \gamma \gamma$ is deeply suppressed by $\eta^4$, with an annihilation cross section about $\pi \alpha^2 \eta^4 / 2 m_\chi^2$ in the nonrelativistic limit. Thus, $\gamma$ lines in this annihilation is far below constraints from the CMB observation \cite{Slatyer:2015jla} and the 21-cm absorption \cite{DAmico:2018sxd}. In this paper, the p-wave annihilation $\bar{\chi} \chi$ $\to A A$ is dominant during millicharged DM freeze-out.

In addition, the millicharged DM of concern can be produced in neutral $\pi^0$'s decay. The transition of $\pi^0 \to \bar{\chi} \chi$ mediated by $A$ is taken in the form
\begin{eqnarray}
T_{\pi \bar{\chi} \chi} \simeq - \frac{\lambda (\xi_u m_u - \xi_d m_d) / v}{\sqrt{2} (m_{\pi^0}^2 -  m_{A}^2)} \frac{f_\pi m_{\pi^0}^2}{m_u + m_d} \pi^0 \bar{\chi} \gamma_5 \chi   ,
\end{eqnarray}
and the decay width $\Gamma_{\bar{\chi} \chi}$ is
\begin{eqnarray} \label{d-width}
\Gamma_{\bar{\chi} \chi} \simeq \frac{\lambda^2 (\xi_u m_u - \xi_d m_d)^2}{16 \pi v^2 (m_{\pi^0}^2 -  m_{A}^2)^2} \frac{f_\pi^2 m_{\pi^0}^5 (1 - \frac{4 m_\chi^2}{ m_{\pi^0}^2} )^{1/2} }{(m_u + m_d)^2}    .
\end{eqnarray}
In SM, the process $\pi^0 \to v \bar{v}$ is forbidden for massless neutrinos \cite{Kalogeropoulos:1979wv,Herczeg:1981xa,Kalloniatis:2005kc}. Thus, the decay $\pi^0 \to \bar{\chi} \chi$ could be employed to search for the millicharged DM.

\section{Numerical analysis}

\begin{figure}[htbp!]
\includegraphics[width=0.4\textwidth]{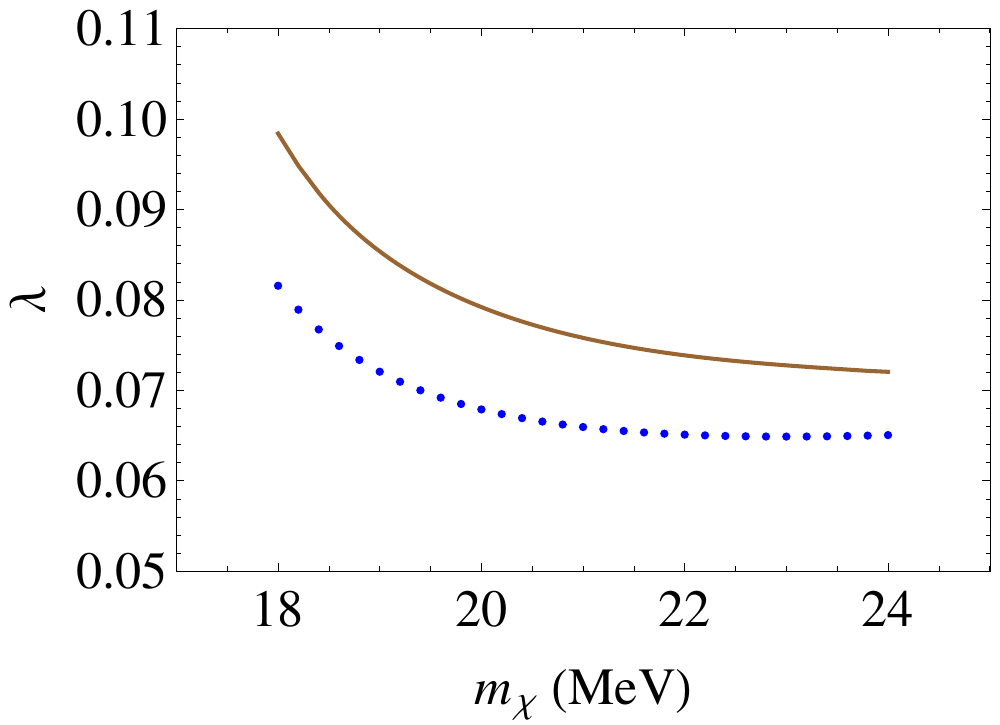} \vspace*{-1ex}
\caption{The value of $\lambda$ for fermionic millicharged DM with the $f_\mathrm{DM}$ required by the EDGES observation. Here $m_A$ = 17 MeV is taken, and the mass range of millicharged DM 18 $ \lesssim m_\chi \lesssim$ 24 MeV is considered. The dotted, solid curves are for the case of $f_\mathrm{DM} =$ $0.4\%$, [$m_{\chi}$ (MeV)/10]$\times$0.115\%, respectively.}\label{dm-lambda}
\end{figure}

The total relic abundance of DM is $\Omega_D h^2 =$ 0.1197 $\pm$ 0.0042 \cite{Ade:2015xua}. The fraction of millicharged DM $f_\mathrm{DM}$ indicated by the EDGES observation is small, about [$m_\chi$ (MeV)/10]$\times$0.115\%$-$0.4\%, and here the mass range of millicharged DM is $2 m_A < 2 m_{\chi} < 3 m_A$. The p-wave process $\bar{\chi} \chi$ $\to A A$ is dominant during millicharged DM freeze-out. To obtain the required fraction $f_\mathrm{DM}$ of millicharged DM, the corresponding coupling parameter $\lambda$ is shown in Fig. \ref{dm-lambda}, with $f_\mathrm{DM} =$ $0.4\%$, [$m_{\chi}$ (MeV)/10]$\times$0.115\%, respectively.

\begin{figure}[htbp!]
\includegraphics[width=0.4\textwidth]{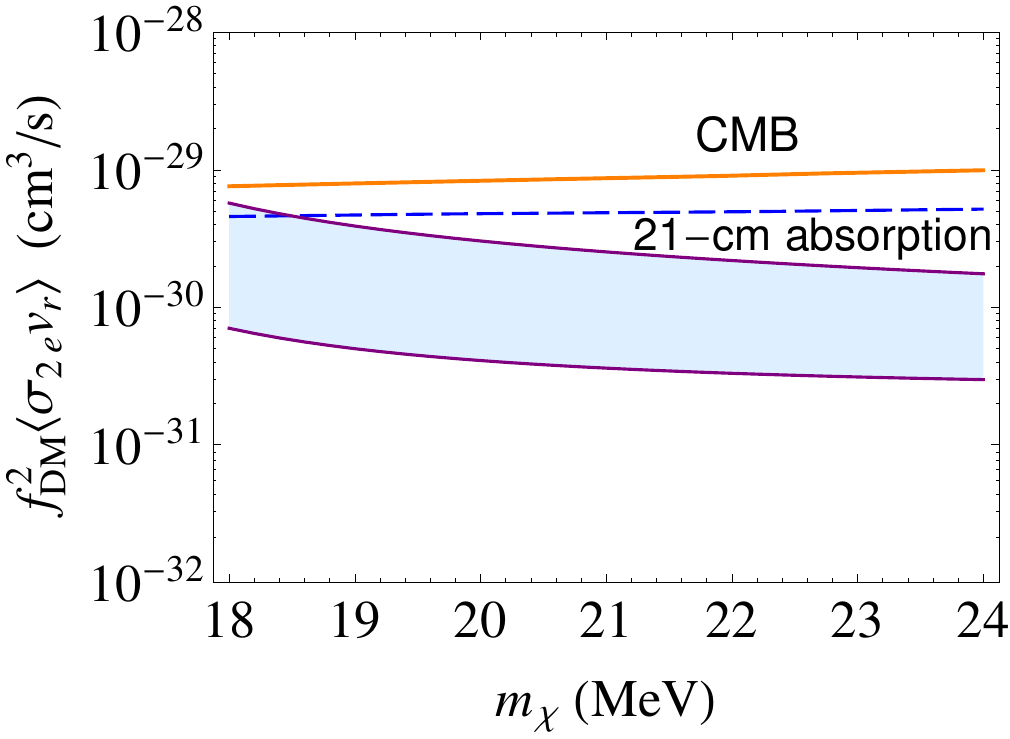} \vspace*{-1ex}
\caption{The range of $f_\mathrm{DM}^2 \langle \sigma_{2e} v_r \rangle$ as a function of $m_{\chi}$. The band is the range of $f_\mathrm{DM}^2 \langle \sigma_{2e} v_r \rangle$, which is set by the limits [$m_{\chi}$ (MeV)/10]$\times$0.115\% $\lesssim f_\mathrm{DM} \lesssim$ $0.4\%$, 115 $\lesssim \xi_e \lesssim 196$ and $10^{-6} \lesssim \eta$ $\lesssim 10^{-4}$. The solid and dashed curves are corresponding to the constraints from the CMB observation \cite{Slatyer:2015jla} and the 21-cm absorption profile (the result noted by Delayed deposition and Boost 1 given by Ref. \cite{DAmico:2018sxd}), respectively.}\label{dm-an}
\end{figure}

The CMB observation \cite{Ade:2015xua,Slatyer:2015jla} and the 21-cm absorption profile \cite{Liu:2018uzy,DAmico:2018sxd} set constraints on the s-wave annihilations $\bar{\chi} \chi$ $\to A, \gamma $ $\to e^+ e^-$. Note an annihilation cross section $\sigma_{2e} v_r \equiv (\sigma_1 + \sigma_2) v_r$ for the annihilation mode $\bar{\chi} \chi$ $\to e^+ e^-$. Considering the limits [$m_{\chi}$ (MeV)/10]$\times$0.115\% $\lesssim f_\mathrm{DM} \lesssim$ $0.4\%$, 115 $\lesssim \xi_e \lesssim$ 196 and $10^{-6} \lesssim \eta$ $\lesssim 10^{-4}$, the range of the weighted annihilation cross section $f_\mathrm{DM}^2 \langle \sigma_{2e} v_r \rangle$ and the constraints are shown in Fig. \ref{dm-an}. It can be seen that, the upper limit of $f_\mathrm{DM}^2 \langle \sigma_{2e} v_r \rangle$ (corresponding to the case of $f_\mathrm{DM} \sim$ $0.4\%$, $\xi_e \sim 196$ and $\eta$ $\sim 10^{-4}$) is allowed by constraints from the CMB observation \cite{Slatyer:2015jla} and the 21-cm absorption profile \cite{DAmico:2018sxd}.

Here we give a brief discussion about the search of millicharged DM at underground experiments. For MeV scale millicharged DM, considering the terrestrial effect of a charged particle penetrating the earth, the exclusion regions of XENON10 \cite{Essig:2012yx,Essig:2017kqs} and COHERENT \cite{Ge:2017mcq} are sensitive for DM with a millicharge parameter $\eta$ $\lesssim 10^{-7}$ \cite{Barkana:2018qrx,Emken:2017erx}. In addition, the millicharged DM residing in the Galactic disk is rare, which is prevented by the magnetic fields in the Milky Way \cite{Barkana:2018lgd,Chuzhoy:2008zy,McDermott:2010pa}. Thus, the millicharged DM of concern is tolerant by the DM direct detections.

Now, we turn to the search of millicharged DM in $\pi^0$'s invisible decay. Some parameters are inputted as follows: $m_{\pi^0} = 134.9766 \pm 0.0006$ MeV, $f_\pi$ = 130.2 (1.7) MeV, i.e., the averaged values from Particle Data Group \cite{Patrignani:2016xqp}. Substituting the corresponding values into Eq. (\ref{d-width}), the decay width $\Gamma_{\bar{\chi} \chi}$ is about
\begin{eqnarray}
\Gamma_{\bar{\chi} \chi} \approx 3.3 \times 10^{-12} (\frac{\lambda}{0.1})^2 (\frac{\xi_u m_u - \xi_d m_d}{m_d})^2  ,
\end{eqnarray}
which is in units of GeV. The mean lifetime of $\pi^0$ is $\tau_{\pi^0} = (8.52 \pm 0.18) \times 10^{-17}$ s \cite{Patrignani:2016xqp}, and thus the branching ratio of the invisible decay $\pi^0 \to \bar{\chi} \chi$ is
\begin{eqnarray}
\mathcal{B}_{\pi^0 \to \bar{\chi} \chi} \approx 4.3 \times 10^{-4} (\frac{\lambda}{0.1})^2 (\frac{\xi_u m_u - \xi_d m_d}{m_d})^2 .
\end{eqnarray}
Experimentally, signatures of the decay $\pi^0 \to$ "invisible" can be searched via the process $K^+ \to \pi^+ \pi^0$ with $\pi^0 \to$ "invisible", which could be identified in kinematics (see e.g., Refs. \cite{Artamonov:2005cu,Moulson:2013oga} for more). The upper limit of $\pi^0$'s invisible decay given by E949 experiment indicates the branching ratio of $\pi^0 \to$ "invisible" $< 2.7 \times 10^{-7}$ \cite{Artamonov:2005cu}. For the case of Higgs-like couplings of $A$ to quarks, i.e., $\xi_u = \xi_d \approx 0.3$, the decay mode $\pi^0 \to \bar{\chi} \chi$ will exceed the upper limit set by the experiment, and thus this case is excluded. For the case of universal couplings of $A$ to up and down quarks, i.e., $m_u \xi_u \sim m_d \xi_d$, the branching ratio $\mathcal{B}_{\pi^0 \to \bar{\chi} \chi}$ will be reduced. Specifically, for $(m_u \xi_u  -  m_d \xi_d)/ m_d \xi_d <$ 0.12, the invisible decay $\pi^0 \to \bar{\chi} \chi$ will be allowed by the upper limit from E949. The millicharged DM of concern can be explored at future high intensity $K^+$ sources, such as NA62. With $\sim 10^{13}$ $K^+$ decays being collected, NA62 would reach a limit of $\sim 10^{-9}$ \cite{Moulson:2013oga} for the branching ratio of $\pi^0 \to$ "invisible".

\section{Conclusion and discussion}

The pseudoscalar $A$ mediated fermionic millicharged DM has been studied in this paper, with $^8$Be anomalous transitions induced by $A$, and contributions from $A$ play the key role in obtaining the small fraction of millicharged DM when millicharged DM freeze-out. The photon mediated scattering between MeV scale millicharged DM and hydrogen could cool the hydrogen and cause the 21-cm absorption at the cosmic dawn. For fermionic millicharged DM $\bar{\chi} \chi$ with the mass in the range of $2 m_A < 2 m_{\chi} < 3 m_A$, the p-wave annihilation $\bar{\chi} \chi \to A A$ could be dominant during DM freeze-out. For the fraction $f_\mathrm{DM}$ of millicharged DM required by the 21-cm absorption, the DM-$A$ coupling parameter $\lambda$ is derived, with $\lambda \sim$ 0.065$-$0.1 for $m_{\chi}$ in a range of 18$-$24 MeV. The s-wave annihilation $\bar{\chi} \chi$ $\to A, \gamma $ $\to e^+ e^-$ is allowed by constraints from CMB and the 21-cm absorption.

The millicharged DM with a millicharge $\eta e$ of concern could evade constraints from direct detection experiments. The $\pi^0$'s invisible decay can be employed to search for the millicharged DM in the process of $K^+ \to \pi^+ \pi^0$ with $\pi^0 \to$ "invisible" (for other approaches, see e.g., Ref. \cite{Gninenko:2014sxa}). For the case of $m_u \xi_u \sim m_d \xi_d$ and $(m_u \xi_u  -  m_d \xi_d)/ m_d \xi_d <$ 0.12, the invisible decay $\pi^0 \to \bar{\chi} \chi$ can be allowed by the upper limit of E949 \cite{Artamonov:2005cu}. With $\mathcal{O}(10^{13})$ $K^+$ decays, NA62 would set a limit of $\sim 10^{-9}$ \cite{Moulson:2013oga} for $\mathcal{B}_{\pi^0 \to \bar{\chi} \chi}$. We look forward to the future accurate 21-cm absorption observations and the run of NA62 experiment, at which the millicharged DM of concern can be tested.

\acknowledgments \vspace*{-3ex} This work was supported by National Natural Science Foundation of China  under the contract No. 11505144, and the Longshan academic talent research supporting program of SWUST
under Contract No. 18LZX415.

\appendix

\section{The annihilation of $\bar{\chi} \chi \to A A A$}
\label{appendix:three-A}

For the mass range of $3 m_A < 2 m_{\chi} < $ 70 MeV, the s-wave annihilation $\bar{\chi} \chi \to A A A$ is opened. Consider the transition$-I$ via the $\chi-A$ coupling $\lambda A \bar{\chi} i \gamma_5 \chi$ at first. Supposing the momentum relation $\chi (p_1) \bar{\chi}(p_2) \to A (k_1) A (k_2) A (k_3)$, the corresponding DM annihilation cross section is
\begin{eqnarray}
\sigma_3^I v_r &\simeq& \frac{1}{2} \frac{1 }{16  m_\chi^2}\frac{1}{3!} \int \mathrm{d}\Phi_3   \\
&& \times | \mathcal{M}_{1,2,3}+ \mathrm{permutations ~ of ~ 1,2,3}|^2, \nonumber
\end{eqnarray}
where $\mathcal{M}_{1,2,3}$ is
\begin{eqnarray}
\mathcal{M}_{1,2,3} = \lambda^3   \frac{ \bar{v}(p_2) \gamma_5 \slashed{k}_3 \slashed{k}_1 u(p_1)}{(m_A^2 - 2 p_2 \cdot k_3)(m_A^2 - 2 p_1 \cdot k_1)} . \nonumber
\end{eqnarray}
In the limit of $p_1^\mu / m_\chi \rightarrow (1, \varepsilon)$, we have
\begin{eqnarray}
\sigma_3^I v_r &\approx& \frac{1}{2} \frac{\lambda^6}{3} \int \mathrm{d}\Phi_3  ( a_{12}^2 + a_{13}^2 + a_{23}^2  \\
&& + 2 a_{12} a_{13} + 2 a_{12} a_{23} + 2 a_{13} a_{23}), \nonumber
\end{eqnarray}
where $a_{ij}$ is
\begin{eqnarray}
a_{ij} \simeq  \frac{  k_i \cdot k_j}{[m_A^2 - (p_1 + p_2) \cdot k_i][m_A^2 - (p_1 + p_2) \cdot k_j]} . \nonumber
\end{eqnarray}
Here we give an estimate about this type DM annihilation. For $f_\mathrm{DM} \sim$ $0.4\%$ and $m_{\chi} \sim$ 30 MeV, the value of $f_\mathrm{DM}^2 \langle \sigma_3^I v_r \rangle$ is about $2 \times 10^{-31} \mathrm{cm}^3/\mathrm{s}$, which is below constraints from the CMB observation \cite{Slatyer:2015jla} and the 21-cm absorption profile \cite{DAmico:2018sxd}.

Now consider the quartic term of $A$,
\begin{eqnarray}
\mathcal{L}_A^{i}= - \frac{\lambda'}{4!} A^4 ,
\end{eqnarray}
and the transition$-II$ $ \bar{\chi} \chi \to A^\ast  \to A  A  A$ occurs. Note $\lambda' = \xi \lambda$. For $f_\mathrm{DM} \sim$ $0.4\%$ and $m_{\chi} \sim$ 30 MeV, the weighted annihilation cross section of transition$-II$ $f_\mathrm{DM}^2 \langle \sigma_3^{II} v_r \rangle$ is about $3.7 \xi^2 \times 10^{-31} \mathrm{cm}^3/\mathrm{s}$. In the case of $\xi \gtrsim 6$, the transition$-II$ is dominant in $A A A$ modes, which is greater than or similar to constraints from the CMB observation and the 21-cm absorption.


\begin{thebibliography}{0} \vspace*{-2ex}



\bibitem{Bowman:2018yin}
  J.~D.~Bowman, A.~E.~E.~Rogers, R.~A.~Monsalve, T.~J.~Mozdzen and N.~Mahesh,
  Nature {\bf 555}, no. 7694, 67 (2018).


\bibitem{Barkana:2018lgd}
  R.~Barkana,
  Nature {\bf 555}, no. 7694, 71 (2018)
  [arXiv:1803.06698 [astro-ph.CO]].


\bibitem{Munoz:2018pzp}
  J.~B.~Mu$\tilde{n}$oz and A.~Loeb,
  Nature {\bf 557}, no. 7707, 684 (2018)
  [arXiv:1802.10094 [astro-ph.CO]].


\bibitem{Xu:2018efh}
  W.~L.~Xu, C.~Dvorkin and A.~Chael,
  Phys.\ Rev.\ D {\bf 97}, no. 10, 103530 (2018)
  [arXiv:1802.06788 [astro-ph.CO]].


\bibitem{Fialkov:2018xre}
  A.~Fialkov, R.~Barkana and A.~Cohen,
  Phys.\ Rev.\ Lett.\  {\bf 121}, 011101 (2018)
  [arXiv:1802.10577 [astro-ph.CO]].


\bibitem{Berlin:2018sjs}
  A.~Berlin, D.~Hooper, G.~Krnjaic and S.~D.~McDermott,
  Phys.\ Rev.\ Lett.\  {\bf 121}, no. 1, 011102 (2018)
  [arXiv:1803.02804 [hep-ph]].


\bibitem{Barkana:2018qrx}
  R.~Barkana, N.~J.~Outmezguine, D.~Redigolo and T.~Volansky,
  arXiv:1803.03091 [hep-ph].


\bibitem{Slatyer:2018aqg}
  T.~R.~Slatyer and C.~L.~Wu,
  Phys.\ Rev.\ D {\bf 98}, no. 2, 023013 (2018)
  [arXiv:1803.09734 [astro-ph.CO]].


\bibitem{Munoz:2018jwq}
  J.~B.~Mu$\tilde{n}$oz, C.~Dvorkin and A.~Loeb,
  Phys.\ Rev.\ Lett.\  {\bf 121}, no. 12, 121301 (2018)
  [arXiv:1804.01092 [astro-ph.CO]].


\bibitem{Mahdawi:2018euy}
  M.~S.~Mahdawi and G.~R.~Farrar,
  arXiv:1804.03073 [hep-ph].


\bibitem{Boddy:2018wzy}
  K.~K.~Boddy, V.~Gluscevic, V.~Poulin, E.~D.~Kovetz, M.~Kamionkowski and R.~Barkana,
  arXiv:1808.00001 [astro-ph.CO].


\bibitem{Kovetz:2018zan}
  E.~D.~Kovetz, V.~Poulin, V.~Gluscevic, K.~K.~Boddy, R.~Barkana and M.~Kamionkowski,
  arXiv:1807.11482 [astro-ph.CO].



\bibitem{Feng:2018rje}
  C.~Feng and G.~Holder,
  Astrophys.\ J.\  {\bf 858}, no. 2, L17 (2018)
  [arXiv:1802.07432 [astro-ph.CO]].


\bibitem{Fraser:2018acy}
  S.~Fraser {\it et al.},
  Phys.\ Lett.\ B {\bf 785}, 159 (2018)
  [arXiv:1803.03245 [hep-ph]].


\bibitem{Mirocha:2018cih}
  J.~Mirocha and S.~R.~Furlanetto,
  arXiv:1803.03272 [astro-ph.GA].


\bibitem{Pospelov:2018kdh}
  M.~Pospelov, J.~Pradler, J.~T.~Ruderman and A.~Urbano,
  Phys.\ Rev.\ Lett.\  {\bf 121}, no. 3, 031103 (2018)
  [arXiv:1803.07048 [hep-ph]].


\bibitem{Moroi:2018vci}
  T.~Moroi, K.~Nakayama and Y.~Tang,
  Phys.\ Lett.\ B {\bf 783}, 301 (2018)
  [arXiv:1804.10378 [hep-ph]].


\bibitem{Barkana:2016nyr}
  R.~Barkana,
  Phys.\ Rept.\  {\bf 645}, 1 (2016)
  [arXiv:1605.04357 [astro-ph.CO]].


\bibitem{Hektor:2018qqw}
  A.~Hektor, G.~H$\ddot{u}$tsi, L.~Marzola, M.~Raidal, V.~Vaskonen and H.~Veerm$\ddot{a}$e,
  Phys.\ Rev.\ D {\bf 98}, no. 2, 023503 (2018)
  [arXiv:1803.09697 [astro-ph.CO]].


\bibitem{Li:2018kzs}
  C.~Li and Y.~F.~Cai,
  arXiv:1804.04816 [astro-ph.CO].


\bibitem{Hektor:2018lec}
  A.~Hektor, G.~H$\ddot{u}$tsi, L.~Marzola and V.~Vaskonen,
  Phys.\ Lett.\ B {\bf 785}, 429 (2018)
  [arXiv:1805.09319 [hep-ph]].


\bibitem{Ade:2015xua}
  P.~A.~R.~Ade {\it et al.} [Planck Collaboration],
  Astron.\ Astrophys.\  {\bf 594}, A13 (2016)
  [arXiv:1502.01589 [astro-ph.CO]].


\bibitem{Slatyer:2015jla}
  T.~R.~Slatyer,
  Phys.\ Rev.\ D {\bf 93}, no. 2, 023527 (2016)
  [arXiv:1506.03811 [hep-ph]].


\bibitem{Cheung:2018vww}
  K.~Cheung, J.~L.~Kuo, K.~W.~Ng and Y.~L.~S.~Tsai,
  arXiv:1803.09398 [astro-ph.CO].


\bibitem{Liu:2018uzy}
  H.~Liu and T.~R.~Slatyer,
  Phys.\ Rev.\ D {\bf 98}, no. 2, 023501 (2018)
  [arXiv:1803.09739 [astro-ph.CO]].


\bibitem{DAmico:2018sxd}
  G.~D'Amico, P.~Panci and A.~Strumia,
  Phys.\ Rev.\ Lett.\  {\bf 121}, no. 1, 011103 (2018)
  [arXiv:1803.03629 [astro-ph.CO]].


\bibitem{Jia:2018csj}
  L.~B.~Jia,
  arXiv:1804.07934 [hep-ph].


\bibitem{Krasznahorkay:2015iga}
  A.~J.~Krasznahorkay {\it et al.},
  Phys.\ Rev.\ Lett.\  {\bf 116}, no. 4, 042501 (2016)
  [arXiv:1504.01527 [nucl-ex]].


\bibitem{Zhang:2017zap}
  X.~Zhang and G.~A.~Miller,
  Phys.\ Lett.\ B {\bf 773}, 159 (2017)
  [arXiv:1703.04588 [nucl-th]].


\bibitem{Feng:2016jff}
  J.~L.~Feng, B.~Fornal, I.~Galon, S.~Gardner, J.~Smolinsky, T.~M.~P.~Tait and P.~Tanedo,
  Phys.\ Rev.\ Lett.\  {\bf 117}, no. 7, 071803 (2016)
  [arXiv:1604.07411 [hep-ph]].


\bibitem{Feng:2016ysn}
  J.~L.~Feng, B.~Fornal, I.~Galon, S.~Gardner, J.~Smolinsky, T.~M.~P.~Tait and P.~Tanedo,
  Phys.\ Rev.\ D {\bf 95}, no. 3, 035017 (2017)
  [arXiv:1608.03591 [hep-ph]].


\bibitem{Kozaczuk:2016nma}
  J.~Kozaczuk, D.~E.~Morrissey and S.~R.~Stroberg,
  Phys.\ Rev.\ D {\bf 95}, no. 11, 115024 (2017)
  [arXiv:1612.01525 [hep-ph]].


\bibitem{DelleRose:2017xil}
  L.~Delle Rose, S.~Khalil and S.~Moretti,
  Phys.\ Rev.\ D {\bf 96}, no. 11, 115024 (2017)
  [arXiv:1704.03436 [hep-ph]].


\bibitem{Gu:2016ege}
  P.~H.~Gu and X.~G.~He,
  Nucl.\ Phys.\ B {\bf 919}, 209 (2017)
  [arXiv:1606.05171 [hep-ph]].


\bibitem{Liang:2016ffe}
  Y.~Liang, L.~B.~Chen and C.~F.~Qiao,
  Chin.\ Phys.\ C {\bf 41}, no. 6, 063105 (2017)
  [arXiv:1607.08309 [hep-ph]].


\bibitem{Fornal:2017msy}
  B.~Fornal,
  Int.\ J.\ Mod.\ Phys.\ A {\bf 32}, 1730020 (2017)
  [arXiv:1707.09749 [hep-ph]].


\bibitem{Banerjee:2018vgk}
  D.~Banerjee {\it et al.} [NA64 Collaboration],
  Phys.\ Rev.\ Lett.\  {\bf 120}, no. 23, 231802 (2018)
  [arXiv:1803.07748 [hep-ex]].


\bibitem{Jiang:2018uhs}
  J.~Jiang, L.~B.~Chen, Y.~Liang and C.~F.~Qiao,
  Eur.\ Phys.\ J.\ C {\bf 78}, no. 6, 456 (2018).


\bibitem{Jia:2016uxs}
  L.~B.~Jia and X.~Q.~Li,
  Eur.\ Phys.\ J.\ C {\bf 76}, no. 12, 706 (2016)
  [arXiv:1608.05443 [hep-ph]].


\bibitem{Kitahara:2016zyb}
  T.~Kitahara and Y.~Yamamoto,
  Phys.\ Rev.\ D {\bf 95}, no. 1, 015008 (2017)
  [arXiv:1609.01605 [hep-ph]].


\bibitem{Chen:2016tdz}
  C.~S.~Chen, G.~L.~Lin, Y.~H.~Lin and F.~Xu,
  Int.\ J.\ Mod.\ Phys.\ A {\bf 32}, no. 31, 1750178 (2017)
  [arXiv:1609.07198 [hep-ph]].


\bibitem{Jia:2017iyc}
  L.~B.~Jia,
  Eur.\ Phys.\ J.\ C {\bf 78}, no. 2, 112 (2018)
  [arXiv:1710.03906 [hep-ph]].


\bibitem{Seto:2016pks}
  O.~Seto and T.~Shimomura,
  Phys.\ Rev.\ D {\bf 95}, no. 9, 095032 (2017)
  [arXiv:1610.08112 [hep-ph]].


\bibitem{Ellwanger:2016wfe}
  U.~Ellwanger and S.~Moretti,
  JHEP {\bf 1611}, 039 (2016)
  [arXiv:1609.01669 [hep-ph]].


\bibitem{Cheng:2012qr}
  H.~Y.~Cheng and C.~W.~Chiang,
  JHEP {\bf 1207}, 009 (2012)
  [arXiv:1202.1292 [hep-ph]].


\bibitem{Berlin:2018bsc}
  A.~Berlin, N.~Blinov, G.~Krnjaic, P.~Schuster and N.~Toro,
  arXiv:1807.01730 [hep-ph].


\bibitem{Riordan:1987aw}
  E.~M.~Riordan {\it et al.},
  Phys.\ Rev.\ Lett.\  {\bf 59}, 755 (1987).


\bibitem{Hektor:2015zba}
  A.~Hektor, K.~Kannike and L.~Marzola,
  JCAP {\bf 1510}, no. 10, 025 (2015)
  [arXiv:1507.05096 [hep-ph]].


\bibitem{Patrignani:2016xqp}
  C.~Patrignani {\it et al.} [Particle Data Group],
  Chin.\ Phys.\ C {\bf 40}, no. 10, 100001 (2016).


\bibitem{Keshavarzi:2018mgv}
  A.~Keshavarzi, D.~Nomura and T.~Teubner,
  Phys.\ Rev.\ D {\bf 97}, no. 11, 114025 (2018)
  [arXiv:1802.02995 [hep-ph]].


\bibitem{Borsanyi:2017zdw}
  S.~Borsanyi {\it et al.} [Budapest-Marseille-Wuppertal Collaboration],
  Phys.\ Rev.\ Lett.\  {\bf 121}, no. 2, 022002 (2018)
  [arXiv:1711.04980 [hep-lat]].


\bibitem{Blum:2018mom}
  T.~Blum {\it et al.} [RBC and UKQCD Collaborations],
  Phys.\ Rev.\ Lett.\  {\bf 121}, no. 2, 022003 (2018)
  [arXiv:1801.07224 [hep-lat]].


\bibitem{Kalogeropoulos:1979wv}
  T.~Kalogeropoulos, J.~Schechter and J.~W.~F.~Valle,
  Phys.\ Lett.\  {\bf 86B}, 72 (1979).


\bibitem{Herczeg:1981xa}
  P.~Herczeg and C.~M.~Hoffman,
  Phys.\ Lett.\  {\bf 100B}, 347 (1981)
  Erratum: [Phys.\ Lett.\  {\bf 102B}, 445 (1981)].


\bibitem{Kalloniatis:2005kc}
  A.~C.~Kalloniatis, J.~D.~Carroll and B.~Y.~Park,
  Phys.\ Rev.\ D {\bf 71}, 114001 (2005)
  [hep-ph/0501117].




\bibitem{Essig:2012yx}
  R.~Essig, A.~Manalaysay, J.~Mardon, P.~Sorensen and T.~Volansky,
  Phys.\ Rev.\ Lett.\  {\bf 109}, 021301 (2012)
  [arXiv:1206.2644 [astro-ph.CO]].


\bibitem{Essig:2017kqs}
  R.~Essig, T.~Volansky and T.~T.~Yu,
  Phys.\ Rev.\ D {\bf 96}, no. 4, 043017 (2017)
  [arXiv:1703.00910 [hep-ph]].


\bibitem{Ge:2017mcq}
  S.~F.~Ge and I.~M.~Shoemaker,
  arXiv:1710.10889 [hep-ph].


\bibitem{Emken:2017erx}
  T.~Emken, C.~Kouvaris and I.~M.~Shoemaker,
  Phys.\ Rev.\ D {\bf 96}, no. 1, 015018 (2017)
  [arXiv:1702.07750 [hep-ph]].


\bibitem{Chuzhoy:2008zy}
  L.~Chuzhoy and E.~W.~Kolb,
  JCAP {\bf 0907}, 014 (2009)
  [arXiv:0809.0436 [astro-ph]].


\bibitem{McDermott:2010pa}
  S.~D.~McDermott, H.~B.~Yu and K.~M.~Zurek,
  Phys.\ Rev.\ D {\bf 83}, 063509 (2011)
  [arXiv:1011.2907 [hep-ph]].



\bibitem{Artamonov:2005cu}
  A.~V.~Artamonov {\it et al.} [E949 Collaboration],
  Phys.\ Rev.\ D {\bf 72}, 091102 (2005)
  [hep-ex/0506028].


\bibitem{Moulson:2013oga}
  M.~Moulson [NA62 Collaboration],
  PoS KAON {\bf 13}, 013 (2013)
  [arXiv:1306.3361 [hep-ex]].


\bibitem{Gninenko:2014sxa}
  S.~N.~Gninenko,
  Phys.\ Rev.\ D {\bf 91}, no. 1, 015004 (2015)
  [arXiv:1409.2288 [hep-ph]].

















\end{thebibliography}
\end{document}